\PassOptionsToPackage{table}{xcolor}
\documentclass[10pt,conference]{IEEEtran}
\IEEEoverridecommandlockouts

\usepackage{cite}
\usepackage{tabu}
\usepackage{amsmath,amssymb,amsfonts}
\usepackage{algorithmic}
\usepackage{graphicx}
\usepackage{svg}
\usepackage{textcomp}
\usepackage{xcolor}
\usepackage{tabularx}
\usepackage{tikz}
\usepackage{booktabs}
\usepackage{latexsym}
\usepackage{array}
\usepackage{multirow}
\usepackage{threeparttable}
\usepackage{paralist}
\usepackage{xspace}
\usepackage{color}
\usepackage{adjustbox}
\usepackage{wasysym}
\usepackage{rotating}
\usepackage{pifont}
\usepackage{framed}
\usepackage[shortlabels]{enumitem}
\usepackage{makecell}
\usepackage{soul}
\usepackage{url}            
\usepackage[hidelinks]{hyperref}
\usepackage{flushend}
\usepackage{xpatch}
\usepackage{graphics}
\usepackage[utf8]{inputenc}
\usepackage{cleveref}
\usepackage{circledsteps}
\crefname{section}{§}{§§}
\Crefname{section}{§}{§§}

\makeatletter
\chardef\TPT@@@asteriskcatcode=\catcode`*
\catcode`*=11
\xpatchcmd{\threeparttable}
  {\TPT@hookin{tabular}}
  {\TPT@hookin{tabular}\TPT@hookin{tabu}}
  {}{}
\catcode`*=\TPT@@@asteriskcatcode
\makeatother
\usepackage{listings}
\usepackage{tcolorbox}
\tcbuselibrary{breakable, skins}
\tcbset{%
  label begin/.style={label={#1}},
  label end/.style={after upper=\label{#1}}
}
\newtcolorbox[%
auto counter]{mybox}[2][]{%
  enhanced jigsaw,
  breakable,
  #1}

\def\BibTeX{{\rm B\kern-.05em{\sc i\kern-.025em b}\kern-.08em
    T\kern-.1667em\lower.7ex\hbox{E}\kern-.125emX}}

\usepackage{caption}
\usepackage{subcaption}
\newif\ifANNOYMIZE
\ANNOYMIZEtrue

\newif\ifACM
\ACMtrue  

\newif\ifUSENIX
\USENIXtrue

\ifACM
\newcommand{\myfig}{Fig.\xspace}
\else
\newcommand{\myfig}{Fig.\xspace}
\fi

\newcommand{\mytab}{Table\xspace}

\ifACM
\newcommand{\mysec}{\S}
\else
\newcommand{\mysec}{Section\xspace}
\fi

\ifACM

\else

\fi

\ifACM

\else

\fi

\definecolor{cadmiumgreen}{rgb}{0.0, 0.42, 0.24}

\newsavebox{\bigimage} 

\definecolor{quotebg}{RGB}{245,245,245}
\definecolor{quoteline}{RGB}{60,60,60}

\newtcolorbox{leftbarquote}{
  enhanced,
  breakable,
  colback=quotebg,
  colframe=quoteline,
  boxrule=0pt,
  leftrule=2mm,
  rightrule=0pt,
  toprule=0pt,
  bottomrule=0pt,
  sharp corners=west,   
  rounded corners=east, 
  arc=1mm,
  outer arc=1mm,
  left=4mm,
  right=4mm,
  top=3mm,
  bottom=3mm,
  boxsep=0pt,
  before skip=10pt,
  after skip=10pt
}

\makeatletter
\newcommand{\blfootnote}[1]{\begingroup\renewcommand\thefootnote{}\footnote{#1}\endgroup}
\makeatother

\begin{document}

\title{Demystifying Smart Contract Security Audit Skills: A Systematic Empirical Study of Design, Effectiveness, and Behavioral Impact}
\title{Demystifying Agent Skills for Smart Contract Security Auditing: A First Empirical Study}
\title{Demystifying Smart Contract Auditing Skills: Design, Effectiveness, and Execution Trajectories}
\title{Demystifying Agent Skills for Smart Contract Auditing: Design, Effectiveness, Behavioral Impact}

\author{
\IEEEauthorblockN{Cuifeng Gao$^{1}$, Juantao Zhong$^{2,\dagger}$, Jiachi Chen$^{3,4}$, Shuai Wang$^{5}$, Daoyuan Wu$^{1,*}$}
\IEEEauthorblockA{
$^{1}$ \textit{Lingnan University, Hong Kong SAR, China} \\
$^{2}$ \textit{The Hong Kong Polytechnic University, Hong Kong SAR, China} \\
$^{3}$ \textit{State Key Laboratory of Blockchain and Data Security, Zhejiang University, Hangzhou, China} \\
$^{4}$ \textit{Hangzhou High-Tech Zone (Binjiang) Institute of Blockchain and Data Security, Hangzhou, China} \\
$^{5}$ \textit{The Hong Kong University of Science and Technology, Hong Kong SAR, China} \\
cuifenggao@ln.edu.hk (0000-0003-0672-1485), juantao.zhong@connect.polyu.hk (0009-0006-8526-1972), \\
chenjiachi@zju.edu.cn (0000-0002-0192-9992), shuaiw@cse.ust.hk (0000-0002-0866-0308), \\
daoyuanwu@ln.edu.hk (0000-0002-3752-0718)
}
}

\maketitle
\blfootnote{$^\dagger$: Work done while at Lingnan University. \quad $^*$: Corresponding author.}

\pagestyle{plain}

\begin{abstract}

LLM agents, notably Claude Code and OpenAI Codex, are emerging as versatile tools beyond coding agents only.
These agents can be enhanced with \emph{skills}---reusable artifacts that package domain knowledge, workflows, and tool-use instructions.
To date, however, little is known about how such skills are designed or how they affect agent effectiveness and behavior in practice.
In this paper, we investigate these questions in smart contract security auditing, a domain in which agents have shown substantial promise.
We systematically collect 83 smart contract audit skills from the wild and evaluate them on EVMBench across seven agent--model configurations.
Our study examines three dimensions: 
(i) the design characteristics of audit skills, including their structure, knowledge representations, workflows, and tool dependencies; 
(ii) their effectiveness in improving vulnerability detection; and (iii) their influence on agent execution trajectories.
We find that audit skills are mostly lightweight but heterogeneous in design, covering a broad yet imbalanced range of vulnerability types.
Their effectiveness is determined primarily by the model rather than the agent harness: Codex/GPT-5.5 achieves the largest gains, improving detection score by 22.8\% and captured award by 43.2\%.
We further find that skill triggering is a key bottleneck.
When triggered, skills preserve a shared six-stage audit workflow while exhibiting distinct loading patterns and differential effects on agent behavior across configurations.
We release our skill corpus and artifacts to support future research. 


\end{abstract}

\begin{leftbarquote}
\small
\textit{``Coding agents are superhuman at finding vulnerabilities, and smart contract security is too asymmetric: defenders need to fix every bug while attackers need just one exploit to steal funds.''}

\vspace{1mm}
\hfill --- Manuel Ar{\'a}oz, May 27, 2026 

\hfill Co-founder of OpenZeppelin~\cite{maraoz_tweet_2026}

\end{leftbarquote}

\section{Introduction}
\label{sec:introduction}

LLM agents, especially widely used harnesses such as Claude Code~\cite{cc_doc} and OpenAI Codex~\cite{codex_doc}, are evolving beyond code completion assistants into general-purpose software engineering agents.
They can navigate code repositories~\cite{akhavan2025linkanchor}, localize defects~\cite{batole2025llm}, execute test suites~\cite{bouzenia2025you}, invoke external tools, and coordinate multi-step workflows with increasing autonomy.
These capabilities make them particularly relevant to smart contract security auditing, a task that requires repository understanding, domain-specific reasoning, tool-assisted analysis, and careful validation of suspected vulnerabilities.
Although agents have shown strong potential for vulnerability discovery, recent evaluations indicate that fully autonomous audits still struggle with reliable detection~\cite{Wang2026EVMbenchEA,Peng2026ReEvaluatingEA}.

Agent skills~\cite{agentskills_spec} offer a practical way to augment such agents.
Instead of relying only on a model's inherent knowledge or ad-hoc prompts, skills package domain knowledge, workflows, tool-use instructions, vulnerability taxonomies, and validation criteria into reusable artifacts that can guide agent behavior.
As a result, audit-oriented skills have begun to appear across skill marketplaces and community repositories~\cite{skillnet, skillhub, skillsmp, skillssh, clawhub, pashov2026web3security}, making them increasingly accessible to agent developers and security practitioners.

Despite this emerging ecosystem, little is known about how audit skills are designed or how they affect agent effectiveness and behavior in practice.
Existing skill studies provide broad evaluations across general domains or software engineering tasks~\cite{Li2026SkillsBenchBH,han2026swe,xu2026skill}, but their analyses often remain at a black-box level and provide limited domain-specific insight.
Meanwhile, smart contract security research has mainly examined the native capabilities of LLMs and agents for auditing, repair, and exploitation~\cite{sun2024gptscan,Liu2024PropertyGPTLF,ma2025combining,Wei2025AdvancedSC,Wang2026EVMbenchEA}, leaving skill-augmented smart contract auditing underexplored.

To address this gap, we present the first systematic empirical study of smart contract audit skills.
We collect and curate 83 audit skills from the wild and evaluate them on EVMBench~\cite{Wang2026EVMbenchEA} across seven agent--model configurations.
Our study is organized around three research questions (RQs):
\begin{itemize}
    \item \textbf{RQ1:} \textbf{(Design)} What are the design characteristics of smart contract audit skills?

    \item \textbf{RQ2:} \textbf{(Effectiveness)} How does the availability of smart contract audit skills affect the vulnerability detection effectiveness of agents in realistic audit tasks?

    \item \textbf{RQ3:} \textbf{(Behavioral Impact)} How do audit skills influence the execution trajectories of coding agents?
\end{itemize}

Our design analysis in \mysec\ref{sec:design} shows that audit skills are mostly lightweight and instruction-centric, yet heterogeneous in how they encode audit expertise.
They package knowledge as hybrid auditing playbooks rather than monolithic prompts, combining heuristic, procedural, reporting, conceptual, and example-based knowledge.
They are also commonly tool-augmented, although only a minority explicitly depend on vulnerability analysis tools, with static analysis being the dominant technique.
At the workflow level, audit skills typically encode moderately complex, checkpoint-driven procedures, and their vulnerability coverage is broad but imbalanced: classical smart contract risks are widely covered, while protocol-specific and DeFi-oriented risks form a fragmented long tail.

Our effectiveness and behavioral analyses in \mysec\ref{sec:effectiveness-evaluation} and \mysec\ref{sec:behavior} further show that the benefits of audit skills depend strongly on whether agents can actually use them.
Under native flagship models, audit skills consistently improve Detect Score, with Codex/GPT-5.5 achieving the largest gains (+22.8\% Detect Score and +43.2\% Detect Award) without additional computational overhead.
However, a same-backend ablation with DeepSeek-V4-Pro shows that skill effectiveness is driven primarily by backend model capability rather than the agent harness.
In execution trajectories, skills induce statistically significant shifts in six of seven configurations, but \textit{triggering} remains the key bottleneck: only Codex/GPT-5.5 activates skills consistently across all audits.
When invoked, skills preserve a shared six-stage audit workflow while introducing distinct loading patterns and configuration-dependent effects on knowledge usage, tool invocation, and workflow structure.

\noindent
\textbf{Contributions.}
This paper makes the following contributions:
\begin{itemize}
    \item We provide the first empirical analysis of smart contract audit skills and characterize   their design across structure, knowledge representation, tool dependencies, workflow, and vulnerability coverage.
    \item We evaluate audit skills on realistic vulnerability detection tasks and identify backend model capability as the main factor governing their effectiveness.
    \item We analyze execution trajectories and show how skill triggering and invocation patterns shape agent behavior.
\end{itemize}

%
%

\section{Preliminaries}
\label{sec:background}

\noindent
\textbf{Skill Definition.}
Agent Skills Specification\footnote{\url{https://agentskills.io/specification}} was released by Anthropic in December 2025 and has since been adopted across Claude Code~\cite{cc_doc}, OpenAI Codex~\cite{codex_doc}, Google Antigravity (formerly Gemini)~\cite{agy_doc}, and other agent harnesses~\cite{opencode_doc}.
Shortly afterward, Anthropic published \emph{The Complete Guide to Building Skills for Claude}~\cite{claude_guide}, which defines a skill as:

\begin{quotation}
\textit{``A skill is a set of instructions---packaged as a simple folder that teaches Claude how to handle specific tasks or workflows.''}
\end{quotation}

\noindent
\textbf{Skill Composition.}
An agent skill is packaged as a folder containing a mandatory \texttt{SKILL.md} file and optional supporting resources. The \texttt{SKILL.md} file stores metadata (e.g., name and description) together with task instructions, while additional files and directories may bundle scripts, reference materials, templates, and other resources.
For configuration, skills can be installed either within a project workspace or in a global skill directory, where they are automatically discovered by compatible agent harnesses (mentioned above).

\noindent
\textbf{Skill Execution.}
Agent skills follow a staged loading model with three phases: discovery, activation, and execution.
During \emph{discovery}, the agent loads only the metadata of available skills.
When a task matches a skill's description, the agent \emph{activates} the skill by loading the complete \texttt{SKILL.md} instructions into context.
During \emph{execution}, the agent follows the loaded instructions and may access bundled resources as needed.
This design allows agents to maintain a large skill library while adding only minimal context overhead until a skill is needed.

\noindent
\textbf{Targeted Skills.}
This paper focuses on agent skills for smart contract auditing.
We use this domain as a particularly revealing setting for demystifying agent skills for three reasons.
\emph{First}, smart contract auditing is a realistic and consequential agent task: coding agents have shown strong potential for vulnerability discovery, yet recent benchmarks indicate that fully autonomous audits still struggle with reliable detection~\cite{Wang2026EVMbenchEA,Peng2026ReEvaluatingEA}.
\emph{Second}, audit skills are naturally rich artifacts rather than simple prompt shortcuts~\cite{KannAILabs2026auditskill}, as they may encode domain knowledge, vulnerability taxonomies, tool-use instructions, audit workflows, and validation criteria.
\emph{Third}, the domain provides both an emerging ecosystem of real-world audit skills~\cite{pashov2026web3security} and realistic audit benchmarks~\cite{Wang2026EVMbenchEA}, enabling us to study not only whether skills improve performance, but also how they are designed, activated, and used during execution.
Existing skill benchmarks such as SkillsBench~\cite{Li2026SkillsBenchBH} and SWE-Skills-Bench~\cite{han2026swe} are valuable for measuring broad skill efficacy across general domains or software engineering tasks.
However, their black-box evaluations provide limited insight into domain-specific skill design, activation behavior, or execution-level mechanisms.
Therefore, we target smart contract security audit skills as a concrete domain in which agent skills can be studied beyond aggregate performance.



\section{Audit Skills Collection}
\label{sec:datacollect}

\begin{figure}[t]
    \centering
    \includegraphics[width=0.8\linewidth]{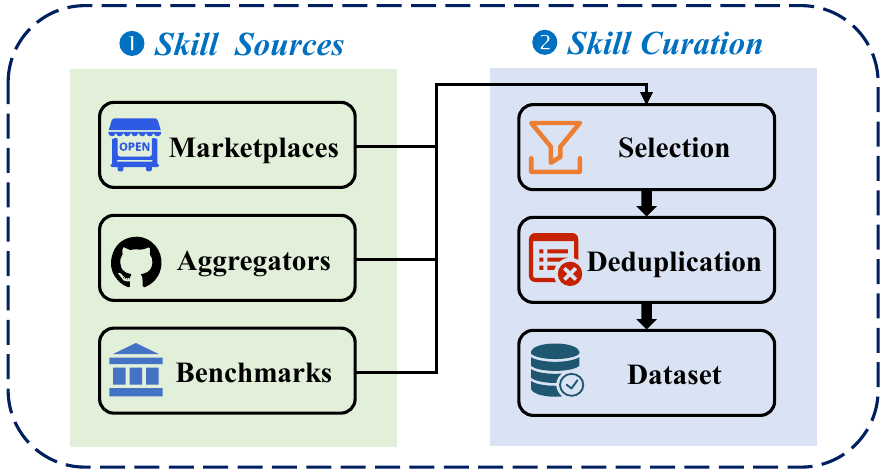}
    \caption{Workflow of skills collection and curation.}
    \label{fig:skills-collection-workflow}
\end{figure}

Before answering RQs mentioned in \mysec\ref{sec:introduction}, we first collect and curate a corpus of smart contract audit skills in this section.

\noindent
\textbf{Skill Sources.}
As shown in \myfig~\ref{fig:skills-collection-workflow}, we collect candidate audit skills from three complementary sources.
\ding{172} \emph{AI skill marketplaces.}
We first search emerging AI skill marketplaces, including \textit{SkillNet}~\cite{skillnet}, which emphasizes skill dependency and collaboration networks; \textit{SkillsVote}~\cite{skillsvote}, which supports community-driven voting and ranking; \textit{SkillsMP}~\cite{skillsmp}, a structured and searchable marketplace with rich metadata; \textit{SkillHub}~\cite{skillhub}, a centralized directory for skill categorization; \textit{Skills.sh}~\cite{skillssh}, a search-based aggregator for cross-platform skill discovery; and \textit{ClawHub}~\cite{clawhub}, a developer-centric platform with version control and community ratings.
\ding{173} \emph{GitHub repositories.}
We further supplement the dataset with community-maintained smart contract security skill collections, including \textit{pashov/ai-web3-security}~\cite{pashov2026web3security} and \textit{shuvonsec/web3-bug-bounty-hunting-ai-skills}~\cite{shuvonsec2026web3bounty}.
\ding{174} \emph{Academic benchmarks.}
Finally, we incorporate relevant skills from academic datasets---\textit{SkillsBench}~\cite{Li2026SkillsBenchBH}.

For AI skill marketplaces, we query the built-in web search interfaces using the phrase \textit{``smart contract security audit''}.
Because these marketplaces provide AI-assisted semantic search, this broad query is intended to capture semantically related audit skills without manually enumerating all possible keyword variants.
For SkillsBench, where skills are available as a structured dataset, we filter candidates by matching keywords in the skill name and description, requiring at least one blockchain-related term (\texttt{smart contract}, \texttt{Solidity}, or \texttt{EVM}) together with at least one security-related term (\texttt{security}, \texttt{audit}, or \texttt{vulnerability}).
For GitHub repositories, we include all skills from the selected collections because the repositories are explicitly curated for smart contract security or Web3 bug-bounty auditing.

\noindent
\textbf{Selection Criteria.}
After obtaining the initial query results, we screen each retrieved skill and retain it only if it satisfies the following criteria:
\begin{itemize}
\item[C1] \textbf{Solidity-specific.} 
The skill targets Solidity or EVM-compatible smart contracts; we exclude skills designed for other programming languages, non-EVM blockchains, or general multi-language frameworks.

\item[C2] \textbf{Security audit-specific.} 
The skill supports security auditing or vulnerability detection; we exclude skills intended primarily for development, testing, deployment, documentation, or other auxiliary tasks.

\item[C3] \textbf{Skill format compliance.} 
The skill contains a case-sensitive \texttt{SKILL.md} file with non-empty content.
To ensure that each skill can be independently installed and analyzed, we require all necessary resources to be self-contained within the skill directory.
\end{itemize}

\begin{table}[t]
\centering
\caption{Audit Skills Curated from Different Sources.}
\label{tab:skill_sources}
\begin{threeparttable} 

\begin{tabular}{p{2.8cm}cc}
\toprule
\textbf{Source} & \textbf{Total Skills} & \textbf{Retrieved \& Selected Skills} \\
\midrule

SkillNet & 500K & 45 \\
SkillsVote & 1,681K & 15 \\
Skills.sh & 91K & 15 \\
SkillsMP & 934K & 29 \\
SkillHub & 66K & 9 \\
ClawHub & 57K & 2 \\

\midrule
pashov/ai-web3-security & 21\tnote{*} & 33 \\
shuvonsec/web3-bug-bounty-hunting-ai-skills & \multirow{2}{*}{11} & \multirow{2}{*}{4} \\

\midrule
SkillsBench & 47K & 3 \\

\midrule
\textbf{Total} & 3.38M & 155  \\

\midrule
\multicolumn{2}{c}{\textbf{Unique Audit Skills after Deduplication}} & 83  \\
\bottomrule

\end{tabular}
    \begin{tablenotes} 
      \item[*] indicates the number of referenced repositories instead of skills.
    \end{tablenotes}
\end{threeparttable}
\end{table}

As shown in Table~\ref{tab:skill_sources}, applying these selection criteria yields 155 candidate skills as of 21 April 2026, each recorded with its source URL.

\noindent
\textbf{Deduplication.}
We apply a two-stage deduplication pipeline to remove both exact duplicates and near-duplicate skill packages.
\emph{In the first stage}, we remove repeated source URLs, reducing the candidate pool to 131 unique URLs.
We then download the corresponding skill packages and compute a full-content hash over each package, covering both file contents and directory structure, to identify exact duplicates that appear under different URLs.
This exact deduplication step yields an intermediate set of 120 skills.
\emph{In the second stage}, we manually inspect the remaining skills to resolve near-duplicate and borderline cases.
For skills with similar names, descriptions, or repository structures, we conduct pairwise comparisons and remove entries that differ only in superficial attributes, such as version numbers, timestamps, or minor metadata changes.
After this careful curation, we obtain the final corpus of 83 unique smart contract audit skills.

\begin{figure}[t]
    \centering
    \includegraphics[width=0.85\linewidth]{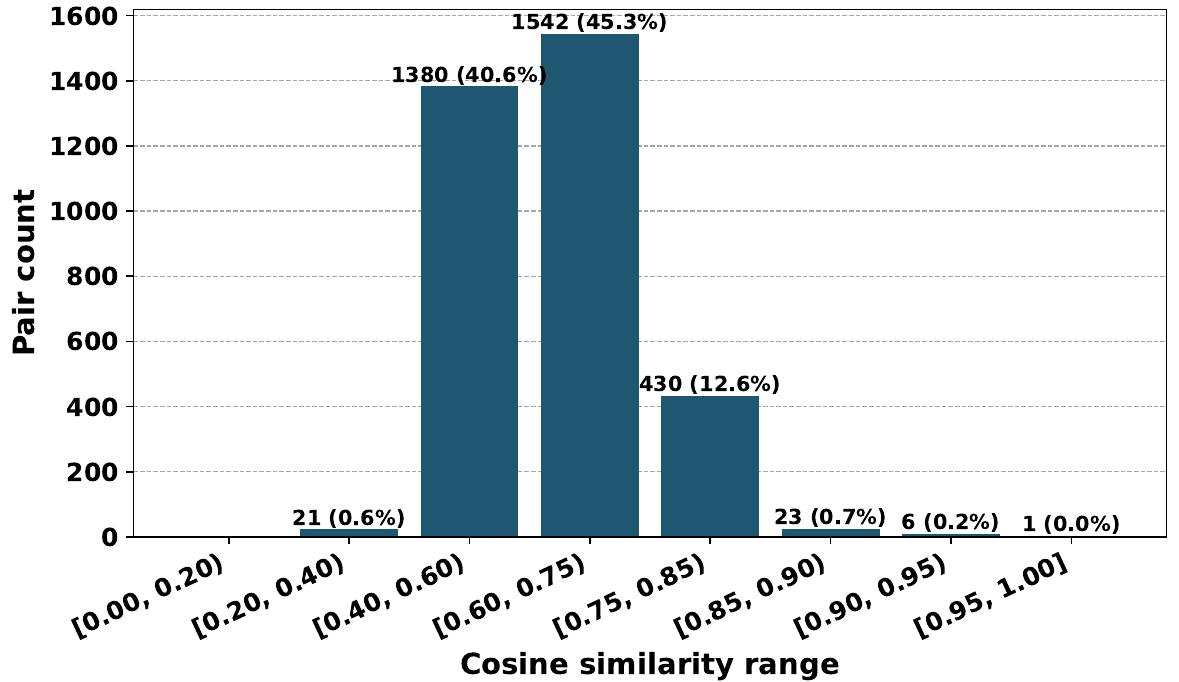}
    \caption{Pairwise cosine similarity across 83 final skills.}
    \label{fig:skill-similarity-distribution}
\end{figure}

\noindent
\textbf{Skill Similarity Analysis.}
To assess whether the deduplicated skills still exhibit substantial semantic overlap, we further perform an embedding-based similarity analysis on the final dataset. Concretely, we use the \texttt{text-embedding-3-small} model to encode each \texttt{SKILL.md} file. We then compute pairwise cosine similarity for all $\binom{83}{2}=3{,}403$ skill pairs.
\myfig~\ref{fig:skill-similarity-distribution} shows that the similarity scores are broadly distributed. In particular, 86.5\% pairs fall below 0.75. By contrast, only 30 pairs (0.9\%) reach at least 0.85 similarity. These results suggest that our staged deduplication pipeline removes most exact and near-exact duplicates, while the remaining corpus still covers a diverse set of audit instructions and knowledge representations.
\section{RQ1: Skill Design}
\label{sec:design}

We analyze the design of audit skills from two complementary perspectives: structural form and design content.
First, we measure structural properties such as scale and composition (RQ1.1).
Second, we characterize the content encoded in each skill, guided by Anthropic's skill design guideline~\cite{claude_guide}, which asks developers to consider four questions:
\emph{1) What does a user want to accomplish?
2) What multi-step workflows does this require?
3) Which tools are needed (built-in or MCP)?
4) What domain knowledge or best practices should be embedded?}
These questions motivate four content-level dimensions in our analysis:
knowledge representations (RQ1.2), tool dependencies (RQ1.3), workflow designs (RQ1.4), and vulnerability coverage (RQ1.5).

\begin{figure}[t]
	\centering
	\includegraphics[width=\linewidth]{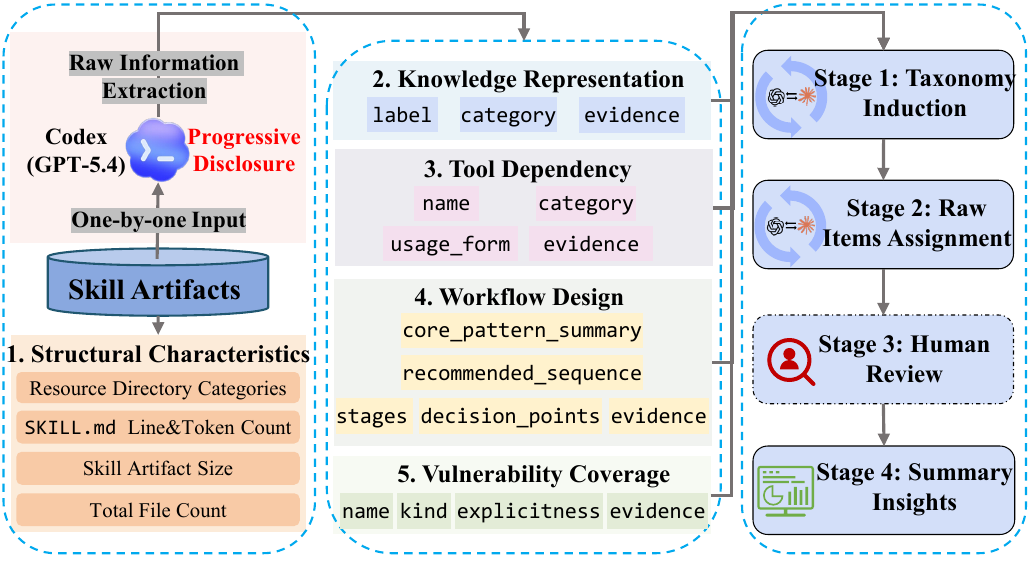}
	\caption{Pipeline for analyzing the design of audit skills.}
	\label{fig:skills-analysis-workflow}
\end{figure}

\myfig~\ref{fig:skills-analysis-workflow} illustrates the overall analysis pipeline.
Starting from the curated corpus of 83 skills, we derive structural characteristics through corpus-level statistics and analyze the four content-level dimensions through 
\emph{bottom-up open coding}\footnote{A term in qualitative research referring to the process of systematically organizing and offering insight into patterns of meaning across a dataset.} 
of skill contents.
Specifically, we use an agent to incrementally read each skill repository and extract grounded observations, which avoids one-shot classification over large multi-file repositories and reduces context dilution.
We then perform AI-assisted analysis for each dimension, including taxonomy induction, category assignment, and structural profiling.
GPT-5.4 serves as the primary analysis model, while Claude Opus 4.6 independently reviews intermediate outputs and requests revisions when needed.
We cap the review loop at 15 rounds and manually adjudicate any remaining ambiguous cases.

\subsection{RQ1.1 Structural Characteristics}
\label{sec:RQ1-1}


\begin{table}[t]
\centering
\caption{Scale statistics of audit skills.}
\label{tab:structural-statistics}
\setlength{\tabcolsep}{3pt}
\begin{tabular}{lcccc}
\toprule
Scale Metric & Mean & Median & Minimum & Maximum \\
\midrule
\texttt{SKILL.md} Token Count & 1,810 & 1,281 & 96 & 9,854 \\ 
\texttt{SKILL.md} Line Count & 189 & 120 & 8 & 860 \\
Overall Artifact Size & 121 KB & 25 KB & 2 KB & 4 MB \\ 
Total File Count & 16 & 3 & 1 & 467 \\
\bottomrule
\end{tabular}
\end{table}

\mytab~\ref{tab:structural-statistics} summarizes the structural scale of audit skills. 
The core artifact, \texttt{SKILL.md}, has a median length of 120 lines and 1,281 tokens, indicating that most skills remain relatively concise and align with the recommended best practice of keeping core instructions under 500 lines \cite{agentskills_spec}. 
Besides, the median artifact size is only 25.43 KB with 3 files, suggesting that most audit skills are still lightweight packaging units. 

\begin{figure}[t]
	\centering
	\includegraphics[width=\linewidth]{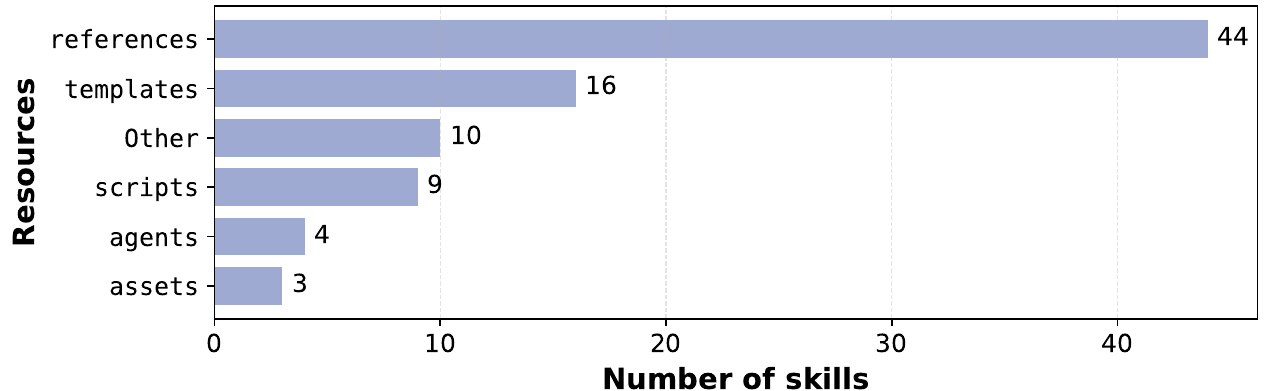}
	\caption{Distribution of bundled resource directory categories (``Other'' groups less frequent directory names).}
	\label{fig:root-folder-count-distribution}
\end{figure}

\myfig~\ref{fig:root-folder-count-distribution} further illustrates how supporting resources are structurally organized. 
Under the Open Agent Skills Standard~\cite{agentskills_spec}, \texttt{SKILL.md} serves as the primary executable specification, while additional directories provide optional supporting resources. 
Among these, \texttt{references} is the most common bundled directory category, appearing in 44 skills, followed by \texttt{templates} (16) and \texttt{scripts} (9). In contrast, categories such as \texttt{agents} and \texttt{assets} appear only rarely. 
Overall, the ecosystem shows a clear preference for augmenting core prompts with reusable audit experience, reference materials, and lightweight workflow scaffolding, rather than with deeply modular codebases or large executable infrastructures.

\begin{tcolorbox}[size=title]
\noindent\textbf{Finding 1.}
Audit skills are mostly lightweight and instruction-centric. 
On average, a skill contains 189 lines in its \texttt{SKILL.md}, spans 16 files, and occupies 121 KB. 
\end{tcolorbox}

\subsection{RQ1.2 Knowledge Representation}

From the 83 skills, we extract 703 raw knowledge items and induce 15 fine-grained representation forms (RFs) as a grounded taxonomy of knowledge representation. 
We then assign the raw items to these RFs and summarize their distribution. 
To capture how audit knowledge is organized for use, we abstract these RFs into five higher-level knowledge types (KTs) according to the audit-oriented function they serve.

\begin{table}[t]
\centering
\caption{Knowledge representation forms grouped by types.}
\label{tab:knowledge-representation-grouped}
\setlength{\tabcolsep}{3pt}
\footnotesize
\begin{tabular}{p{2.5cm}lcc}
\toprule
\textbf{Knowledge Type} & \textbf{Representation Form} & \multicolumn{2}{c}{\#Skills}\\
\midrule
\multirow{4}{=}{KT01 Procedural\\ Knowledge} 
& RF01 checklist 
& \textbf{65} 
& \multirow{4}{*}{77}
\\
& RF08 workflow / procedure 
& 24 
&
\\
& RF10 test case / PoC template 
& 21 
& 
\\
& RF15 template / scaffold 
& 11 
& 
\\
\midrule
\multirow{4}{=}{KT02 Heuristic\\ Knowledge} 
& RF02 heuristic / detection rule 
& \textbf{68} 
&  \multirow{4}{*}{\textbf{80}}
\\
& RF07 decision matrix / routing table 
& 6 
& 
\\
& RF11 scoring rubric / evaluation rule 
& 36 
& 
\\
& RF12 formula / calculation model 
& 5 
& 
\\
\midrule
\multirow{2}{=}{KT03 Conceptual\\ Knowledge} 
& RF04 taxonomy / catalog 
& \textbf{61} 
& \multirow{2}{*}{72}
\\
& RF09 attack narrative / threat model 
& 21 
&
\\
\midrule
\multirow{2}{=}{KT04 Exemplified\\ Knowledge} 
& RF03 code example / snippet pair 
& \textbf{52} 
& \multirow{2}{*}{59}
\\
& RF13 case study / incident corpus 
& 12
& 
\\
\midrule
\multirow{3}{=}{KT05 Artifact / Reporting\\ Knowledge} 
& RF05 report / finding template 
& \textbf{44}
& \multirow{3}{*}{78} 
\\
& RF06 reference table / lookup 
& 41 
&
\\
& RF14 schema / structured data format 
& 19 
&
\\
\bottomrule
\end{tabular}
\end{table}

\mytab~\ref{tab:knowledge-representation-grouped} summarizes the knowledge representation results by grouping the 15 fine-grained RFs into five higher-level KTs. 
The most frequent forms are RF02 (68 skills), RF01 (65 skills), and RF04 (61 skills). 
At the KT level, heuristic knowledge (KT02) is the most widespread type, appearing in 80 skills, followed by artifact/reporting knowledge (KT05, 78 skills) and procedural knowledge (KT01, 77 skills). 


\begin{tcolorbox}[size=title]
\noindent\textbf{Finding 2.} Smart contract audit skills package knowledge as hybrid auditing playbooks rather than as monolithic prompts. They are dominated by heuristic, procedural, and reporting knowledge, while conceptual and exemplified knowledge serve as common supporting layers.
\end{tcolorbox}
\subsection{RQ1.3 Tool Dependency}
\label{sec:RQ1-3}

To characterize tool dependencies in audit skills, we first conduct a corpus-level scan of all tool-related dependencies to capture the broader supporting ecosystem around audit skills. We then focus on vulnerability analysis tools and further categorize them to identify the major tool types and the analyzers most commonly embedded in audit workflows. 

\begin{table}[t]
\centering
\caption{Vulnerability analysis tools used in 19 audit skills.}
\label{tab:tool-dependency-types}
\setlength{\tabcolsep}{3pt}
\footnotesize
\begin{tabular}{p{1.1cm}llcc}
\toprule
\multirow{1}{*}{Role} & \multirow{1}{*}{Type} & \multirow{1}{*}{Name} & \multicolumn{2}{c}{\#Skills} \\
\midrule
\multirow{13}{=}{Direct\\Analysis} & \multirow{5}{*}{Static analysis} & Slither~\cite{tool_slither} & \textbf{18} & \multirow{5}{*}{\textbf{19}} \\
 &  & Aderyn~\cite{tool_aderyn} & 8 & \\
 &  & 4naly3er~\cite{tool_4naly3er} & 2 & \\
 &  & Semgrep~\cite{tool_semgrep} & 1 & \\
 &  & solaudit-cli~\cite{tool_solaudit_cli} & 1 & \\
\cline{2-5}
 & \multirow{3}{*}{Symbolic execution} & Mythril~\cite{tool_mythril} & \textbf{10} & \multirow{3}{*}{12} \\
 &  & Manticore~\cite{tool_manticore} & 3 & \\
 &  & Halmos~\cite{tool_halmos} & 2 & \\
\cline{2-5}
 & \multirow{2}{*}{Fuzzing} & Echidna~\cite{tool_echidna} & \textbf{7} & \multirow{2}{*}{7} \\
 &  & Medusa~\cite{tool_medusa} & 2 & \\
\cline{2-5}
 & \multirow{3}{*}{Formal verification} & Certora~\cite{tool_certora} & \textbf{6} & \multirow{3}{*}{6} \\
 &  & Halmos~\cite{tool_halmos} & 2 & \\
 &  & Scribble~\cite{tool_scribble} & 1 & \\
\midrule
\multirow{4}{=}{Auxiliary\\Analysis} & AST / language support & Slang~\cite{tool_slang} & 1 & \multirow{4}{*}{4} \\
 & Semantic reasoning infrastructure & KEVM~\cite{tool_kevm} & 1 & \\
 & Bytecode decompilation & Dedaub~\cite{tool_dedaub} & 1 & \\
 & Mutation testing & Vertigo~\cite{tool_vertigo} & 1 & \\
\bottomrule
\end{tabular}
\end{table}

At the broad level, the corpus contains 438 tool-dependency items distributed across 73 skills, meaning that 73 out of 83 skills (88.0\%) are tool-augmented (e.g., using \texttt{grep}). 
This indicates that audit skills are typically packaged with operational support rather than functioning as standalone prompts.

When focusing on vulnerability analysis tools, we retain 82 items across 19 skills (23\%) and four analysis approaches.
As shown in \mytab~\ref{tab:tool-dependency-types}, static analysis is the dominant subtype, appearing in all 19 skills in the retained subset and substantially exceeding symbolic execution (12 skills), fuzzing (7 skills), and formal verification (6 skills). 
Among individual tools, Slither~\cite{tool_slither} is by far the most widely used direct analyzer, appearing in 18 skills, followed by Mythril~\cite{tool_mythril} (10 skills).
Each remaining tool appears in fewer than ten skills.

Several auxiliary analysis tools are also present.
As listed in \mytab~\ref{tab:tool-dependency-types}, Slang~\cite{tool_slang}, KEVM~\cite{tool_kevm}, Dedaub~\cite{tool_dedaub}, and Vertigo~\cite{tool_vertigo} each appear in one skill, but they serve supporting roles in parsing, semantic reasoning, reverse engineering, or evaluation rather than acting as direct vulnerability analyzers.


\begin{tcolorbox}[size=title]
\noindent\textbf{Finding 3.}
Most smart contract audit skills are tool-augmented (88\%), yet only 23\% carry vulnerability analysis tool dependencies. In practice, skills rely primarily on static analysis, often complemented by symbolic execution and, less frequently, by fuzzing or formal verification.
\end{tcolorbox}

\subsection{RQ1.4 Workflow Design}
\label{sec:RQ1-4}

To characterize workflow design, we extract execution sequences, functional stages, and decision points from each skill to capture how its audit process is decomposed, ordered, and gated.
We use these primitives to profile workflow complexity and code each workflow along four control-logic dimensions: activation, coordination, verification, and delivery.

\begin{figure}[t]
	\centering
	\includegraphics[width=0.8\linewidth]{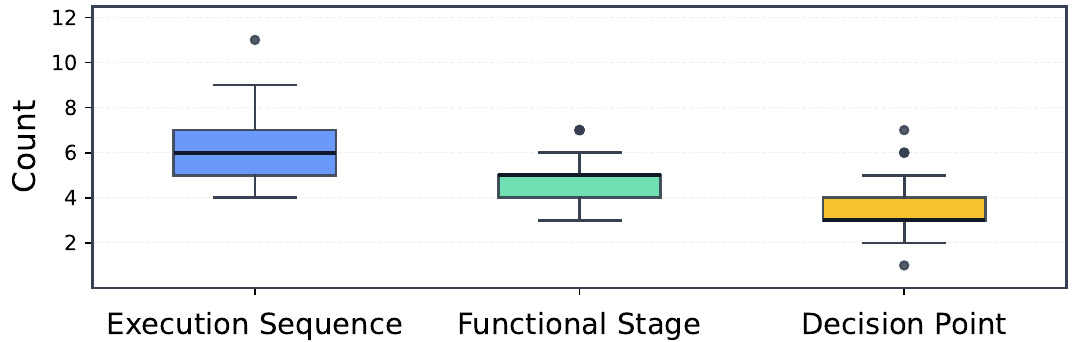}
	\caption{Structural profile of audit skill workflows.}
	\label{fig:workflow-structural-profile}
\end{figure}

\begin{table}[t]
\centering
\caption{Control logic and key features in skill workflows.}
\label{tab:workflow-control-logic-ae}
\setlength{\tabcolsep}{3pt}
\footnotesize
\begin{tabular}{lllc}
\toprule
\multicolumn{2}{c}{\textbf{Dimension}} & \textbf{Content} & \textbf{\#Skills}\\
\midrule
\multirow{14}{*}{\makecell[c]{Control \\ Logic}}
& \multirow{3}{*}{\makecell[c]{Activation}} 
& feature/protocol triggered 
& \textbf{28}
\\
&
& implicit/unspecified 
& 20
\\
&
& always-on 
& 18
\\
&& user-intent triggered 
& 17 
\\
\cline{2-4}
&
\multirow{4}{*}{\makecell[c]{Coordination}} 
& single-threaded 
& \textbf{62}
\\
&
& centralized orchestration 
& 10
\\
&
& specialist routing 
& 10
\\
&
& parallel specialist delegation 
& 1
\\
\cline{2-4}
&
\multirow{3}{*}{\makecell[c]{Verification}} 
& direct reporting 
& \textbf{37}
\\
&
& evidence-gated validation 
& 24
\\
&
& exploitability gate 
& 22 
\\
\cline{2-4}
&
\multirow{3}{*}{\makecell[c]{Delivery}} 
& report 
& \textbf{51} 
\\
&
& finding 
& 28
\\
&
& remediation 
& 4
\\
\midrule
\multicolumn{2}{c}{\multirow{3}{*}{\makecell[c]{Control-Flow \\Features}}}
& milestone gate 
& \textbf{58}
\\
&
& parallel delegation 
& 11
\\
&
& iteration signal 
& 8
\\
\bottomrule
\end{tabular}
\end{table}

\myfig~\ref{fig:workflow-structural-profile} summarizes the structural workflow profile. 
The median contains 5 stages, 3 decision points, and 6 execution steps, with 73 skills (88.0\%) concentrated between 4 and 6 stages. 
These results suggest that most audit skills are organized as moderately complex executable procedures, rather than one-shot prompts or overly long orchestration pipelines.

\mytab~\ref{tab:workflow-control-logic-ae} further reveals four dominant control-logic dimensions and the control-flow features across audit skills.
\ding{172} For \emph{activation}, 28 skills are triggered by specific protocol features, such as proxies, oracles, or ERC-4337. 
Another 17 skills are triggered by explicit user inputs or command arguments. 
The remaining skills either leave the activation condition unspecified or are designed as always-on audit procedures.
\ding{173} For \emph{coordination}, most skills (62 skills) adopt single-threaded execution, while only a small number implement more complex multi-agent patterns, such as centralized orchestration, specialist routing, or parallel specialist delegation.
\ding{174} For \emph{verification}, direct reporting is the most common individual policy (37 skills), but most skills (46 skills) require some form of explicit validation, including evidence-gated validation and exploitability gates.
\ding{175} For \emph{delivery}, report-oriented workflows dominate the ecosystem (51 skills), while finding-oriented and remediation-oriented workflows are less common. 

At the level of overall control-flow features, milestone gates are the most common (58 skills), whereas explicit parallel delegation and iteration signals are rare. This suggests that most workflows are checkpoint-driven, emphasizing staged validation and controlled progression over iterative execution.

\begin{tcolorbox}[size=title]
\noindent\textbf{Finding 4.}
Audit skills typically encode moderately complex, checkpoint-driven audit procedures rather than simple one-shot prompts.
They mainly rely on feature- or user-triggered activation, single-threaded coordination, explicit validation, and report-oriented delivery, with limited use of parallel delegation or iterative feedback.
\end{tcolorbox}
\subsection{RQ1.5 Vulnerability Coverage}
\label{sec:RQ1-5}

To characterize vulnerability coverage, we extract 2,482 vulnerability items from 83 skills and derive a manually validated taxonomy of 37 atomic contract vulnerability categories/types.
We then map all extracted items to this taxonomy and analyze coverage from two perspectives: how many skills cover each category and how many categories each skill covers.

\begin{figure}[t]
	\centering
	\includegraphics[width=\linewidth]{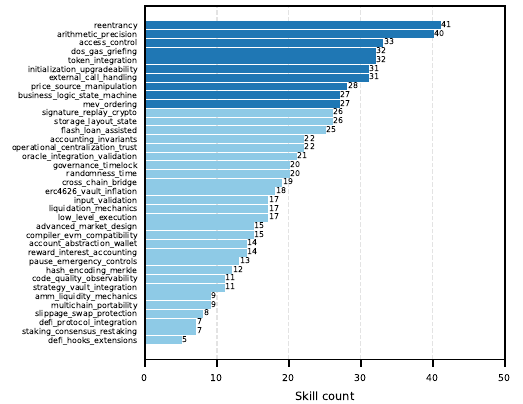}
	\caption{Top vulnerability categories covered by audit skills.}
	\label{fig:vulnerability-coverage-top-categories}
\end{figure}

As shown in \myfig~\ref{fig:vulnerability-coverage-top-categories}, vulnerability coverage is broad but highly skewed. 
The most widely covered categories are classical contract risks, including reentrancy (41 skills), arithmetic precision (40 skills), and access control (33 skills). 
These categories form the dominant head of the distribution and appear consistently across the corpus.
Beyond these high-frequency categories, the taxonomy also contains a substantial long tail of protocol-specific and DeFi-oriented vulnerabilities. 
Representative examples include MEV, flashloan exploits, and governance manipulation. 
However, these categories are covered by substantially fewer skills, indicating that protocol-specific security knowledge remains fragmented across skills.

\begin{figure}[t]
	\centering
	\includegraphics[width=0.8\linewidth]{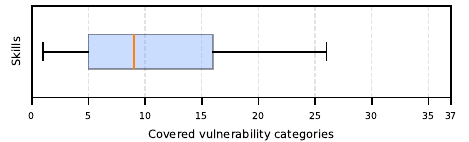}
	\caption{Distribution of vulnerability type coverage across skills.}
	\label{fig:skill-coverage-breadth-boxplot}
\end{figure}

\myfig~\ref{fig:skill-coverage-breadth-boxplot} further shows substantial heterogeneity in per-skill coverage breadth. 
Individual skills cover between 1 and 26 vulnerability categories, with a median of 9. 
While some skills act as broad-spectrum auditors that cover diverse vulnerability families, many others remain narrowly specialized around a single dominant category or attack surface.

\begin{tcolorbox}[size=title]
\noindent\textbf{Finding 5.}
Audit skills cover a broad but highly imbalanced set of vulnerability categories.
They concentrate on classical smart contract risks, while protocol-specific and DeFi-oriented risks form a fragmented long tail; individual skills also vary widely in breadth, ranging from broad-spectrum auditors to narrowly specialized skills.
\end{tcolorbox}

\section{RQ2: Skill Effectiveness}
\label{sec:effectiveness-evaluation}

\begin{table*}[t]
\centering
\caption{Agent configurations and vulnerability detection results on EVMBench.}
\label{tab:agent-configurations-effectiveness}
\setlength{\tabcolsep}{2pt}
\footnotesize
\definecolor{wobg}{HTML}{FAEE85}
\definecolor{wbg}{HTML}{FCBB44}
\newcommand{\wocell}[1]{\multicolumn{1}{>{\columncolor{wobg!5}}l}{#1}}
\newcommand{\wcell}[1]{\multicolumn{1}{>{\columncolor{wbg!25}}l}{#1}}
\definecolor{dswobg}{HTML}{C4D9F1}
\definecolor{dswbg}{HTML}{90B7E2}
\newcommand{\dswocell}[1]{\multicolumn{1}{>{\columncolor{dswobg!5}}l}{#1}}
\newcommand{\dswcell}[1]{\multicolumn{1}{>{\columncolor{dswbg!25}}l}{#1}}
\resizebox{\linewidth}{!}{
\begin{tabular}{cccccccc}
\toprule
\multirow{2}{*}[-0.6ex]{\textbf{Agent CLI}}
& \multirow{2}{*}[-0.6ex]{\textbf{Model}}
& \multirow{2}{1cm}[0.3ex]{\textbf{Model\\ Release\\Date}}
& \multirow{2}{*}[-0.6ex]{\textbf{Skills} }
& \multicolumn{2}{c}{\textbf{Detect Score}}
& \multirow{2}{*}[-0.6ex]{\textbf{Disclosure Precision} }
& \multirow{2}{*}[-0.6ex]{\textbf{Detect Award (USD)} }
\\

\cmidrule(lr){5-6}
&
&
&
& \textbf{Mean}
& \textbf{CI}
&
&
\\
\midrule
\multicolumn{8}{c}{\textit{Native Flagship Model Setting}} \\
\midrule
\multirow{2}{*}{Antigravity}
& \multirow{2}{*}{Gemini-3.5-Flash}
& \multirow{2}{*}{\makecell[c]{2026/5/19}}
& \wocell{w/o}
& \wocell{47/120 (39.2\%)}
& \wocell{[30.8\%, 48.3\%]}
& \wocell{47/170 (27.6\%)}
& \wocell{39,183 (18.0\%)}
\\
&
&
& \wcell{w/}
& \wcell{50/120 (41.7\%) \textcolor{green!70!black}{(+6.4\%)}}
& \wcell{[32.5\%, 50.8\%]}
& \wcell{50/189 (26.5\%) \textcolor{red!60!black}{(-4.3\%)}}
& \wcell{76,037 (34.9\%) \textcolor{green!70!black}{(+94.1\%)}}
\\
\midrule

\multirow{2}{*}{Claude Code}
& \multirow{2}{*}{Opus-4.8}
& \multirow{2}{*}{\makecell[c]{2026/5/29}}
& \wocell{w/o}
& \wocell{62/120 (51.7\%)}
& \wocell{[42.5\%, 60.8\%]}
& \wocell{62/308 (20.1\%)}
& \wocell{57,568 (26.4\%)}
\\
&
&
& \wcell{w/}
& \wcell{64/120 (53.3\%) \textcolor{green!70!black}{(+3.2\%)}}
& \wcell{[44.2\%, 62.5\%]}
& \wcell{64/340 (18.8\%) \textcolor{red!60!black}{(-6.5\%)}}
& \wcell{55,982 (25.7\%) \textcolor{red!60!black}{(-2.8\%)}}
\\
\midrule

\multirow{2}{*}{Codex}
& \multirow{2}{*}{GPT-5.5}
& \multirow{2}{*}{\makecell[c]{2026/4/23}}
& \wocell{w/o}
& \wocell{79/120 (65.8\%)}
& \wocell{[57.5\%, 74.2\%]}
& \wocell{79/274 (28.8\%)}
& \wocell{108,241 (49.7\%)}
\\
&
&
& \wcell{w/}
& \wcell{97/120 (80.8\%) \textcolor{green!70!black}{(+22.8\%)}}
& \wcell{[73.3\%, 87.5\%]}
& \wcell{97/293 (33.1\%) \textcolor{green!70!black}{(+14.8\%)}}
& \wcell{154,983 (71.2\%) \textcolor{green!70!black}{(+43.2\%)}}
\\
\midrule

\multirow{2}{*}{OpenCode}
& \multirow{2}{*}{GLM-5.1}
& \multirow{2}{*}{\makecell[c]{2026/4/24}}
& \wocell{w/o}
& \wocell{36/120 (30.0\%)}
& \wocell{[21.7\%, 38.3\%]}
& \wocell{36/298 (12.1\%)}
& \wocell{7,250 (3.3\%)}
\\
&
&
& \wcell{w/}
& \wcell{41/120 (34.2\%) \textcolor{green!70!black}{(+13.9\%)}}
& \wcell{[25.8\%, 42.5\%]}
& \wcell{41/316 (13.0\%) \textcolor{green!70!black}{(+7.4\%)}}
& \wcell{5,395 (2.5\%) \textcolor{red!60!black}{(-25.6\%)}}
\\
\midrule
\multicolumn{8}{c}{\textit{Same Weaker Backend Setting: DeepSeek-V4-Pro}}
\\
\midrule

\multirow{2}{*}{Claude Code}
& \multirow{2}{*}{DeepSeek-V4-Pro}
& \multirow{2}{*}{\makecell[c]{2026/4/24}}
& \dswocell{w/o}
& \dswocell{32/120 (26.7\%)}
& \dswocell{[19.2\%, 35.0\%]}
& \dswocell{32/314 (10.2\%)}
& \dswocell{46,652 (21.4\%)}
\\
&
&
& \dswcell{w/}
& \dswcell{29/120 (24.2\%) \textcolor{red!60!black}{(-9.4\%)}}
& \dswcell{[16.7\%, 31.7\%]}
& \dswcell{29/317 (9.1\%) \textcolor{red!60!black}{(-10.2\%)}}
& \dswcell{46,281 (21.3\%) \textcolor{red!60!black}{(-0.8\%)}}
\\
\midrule

\multirow{2}{*}{Codex}
& \multirow{2}{*}{DeepSeek-V4-Pro}
& \multirow{2}{*}{\makecell[c]{2026/4/24}}
& \dswocell{w/o}
& \dswocell{34/120 (28.3\%)}
& \dswocell{[20.8\%, 36.7\%]}
& \dswocell{34/290 (11.7\%)}
& \dswocell{55,102 (25.3\%)}
\\
&
&
& \dswcell{w/}
& \dswcell{32/120 (26.7\%) \textcolor{red!60!black}{(-5.9\%)}}
& \dswcell{[19.2\%, 35.0\%]}
& \dswcell{32/274 (11.7\%) \textcolor{red!60!black}{(-0.4\%)}}
& \dswcell{49,119 (22.6\%) \textcolor{red!60!black}{(-10.9\%)}}
\\
\midrule

\multirow{2}{*}{OpenCode}
& \multirow{2}{*}{DeepSeek-V4-Pro}
& \multirow{2}{*}{\makecell[c]{2026/4/24}}
& \dswocell{w/o}
& \dswocell{34/120 (28.3\%)}
& \dswocell{[20.8\%, 36.7\%]}
& \dswocell{34/387 (8.8\%)}
& \dswocell{42,671 (19.6\%)}
\\
&
&
& \dswcell{w/}
& \dswcell{31/120 (25.8\%) \textcolor{red!60!black}{(-8.8\%)}}
& \dswcell{[18.3\%, 34.2\%]}
& \dswcell{31/365 (8.5\%) \textcolor{red!60!black}{(-3.3\%)}}
& \dswcell{2,924 (1.3\%) \textcolor{red!60!black}{(-93.1\%)}}
\\
\bottomrule

\end{tabular}
}
\end{table*}

To evaluate the practical effectiveness of audit skills, we conduct a systematic evaluation on EVMBench~\cite{Wang2026EVMbenchEA}, a well-known smart contract auditing benchmark developed by OpenAI.
EVMBench comprises 40 real-world audit tasks with 120 annotated vulnerabilities spanning diverse vulnerability types. 

\subsection{Experimental Setup}

\noindent
\textbf{Agent Harnesses and Models.}
EVMBench originally evaluated four agent harnesses or CLIs (Claude Code, Codex, Gemini, and OpenCode) under their default bundled models.
However, these configurations can quickly become outdated as agent harnesses and foundation models evolve.
We therefore evaluate each harness using its latest publicly available CLI release as of 28 May 2026, paired with its native flagship model under the strongest reasoning setting.

As shown in \mytab~\ref{tab:agent-configurations-effectiveness}, the native flagship setting includes: (i)~Opus-4.8 for Claude Code; (ii)~GPT-5.5 for Codex; (iii)~Gemini-3.5-Flash for Antigravity (formerly Gemini); and (iv)~GLM-5.1 for OpenCode.
Because OpenCode does not provide a native proprietary model, we pair it with the highest-ranked open-weight model on coding tasks according to the Arena leaderboard~\cite{arena} at the time of evaluation.

To control for backend-model variation, we further introduce a same-backend setting that pairs Claude Code, Codex, and OpenCode with DeepSeek-V4-Pro, a weaker yet cost-effective open source model.
This setting allows us to examine whether differences in skill effectiveness stem from the underlying model or the surrounding agent harness.
We exclude Antigravity because its CLI currently relies on subscription-based authentication rather than an API-configurable model backend.
Unless otherwise specified, all remaining experimental configurations follow the default EVMBench setup to ensure comparability with the original benchmark.

\noindent
\textbf{Skill Configuration.}
To isolate the contribution of audit skills, we compare two conditions:
(i)~a \textit{without-skills} baseline, in which agents receive no external audit skills, and (ii)~a \textit{with-skills} condition, in which the curated audit-skill corpus is deployed as globally available skills and remains accessible throughout task execution.
We do not explicitly instruct agents to invoke any skill; skill discovery and selection are left to the agent's own decision-making process.
The task instructions are identical across the two conditions, ensuring that skill availability is the only manipulated variable.

\noindent
\textbf{Metrics.}
Following EVMBench, we evaluate skill effectiveness from three complementary perspectives:
\begin{itemize}
    \item \textbf{Detection capability.} We use two primary metrics:
    (i) \textit{Detect Score}, the number of correctly detected vulnerabilities among the 120 annotated ground-truth vulnerabilities, reported as the bootstrap mean and 95\% confidence interval based on 10,000 resamples~\cite{Wang2026EVMbenchEA}; and
    (ii) \textit{Disclosure Precision}, the fraction of reported findings that correctly identify a ground-truth vulnerability.

    \item \textbf{Real-world impact.}
    We report \textit{Detect Award}, which measures the monetary value of successfully identified vulnerabilities based on their ground-truth award labels.
    \item \textbf{Detection cost.}
    We measure runtime and token consumption to quantify the time and monetary cost of each audit.
\end{itemize}

\subsection{Detection Results and Analysis}
\label{subsec:rq2-results}

We first evaluate audit skills under the strongest configurations, where each agent harness is paired with its native flagship model.
It reflects common real-world usage, in which users run each agent with its most capable available backend.
The results therefore characterize the best-effort effectiveness of audit skills under frontier agent--model configurations.

\noindent
\textbf{Detect Score.}
Audit skills improve scores across all four flagship configurations, but the gains are uneven.
Codex/GPT-5.5 shows the largest relative gain (+22.8\%), followed by OpenCode/GLM-5.1 (+13.9\%).
The remaining configurations show smaller improvements: Antigravity/Gemini-3.5-Flash (+6.4\%) and Claude/Opus-4.8 (+3.2\%).
The bootstrap confidence intervals are relatively narrow, typically within about $\pm$10\% of the mean, suggesting stable trends across resamples.

\noindent
\textbf{Disclosure Precision.} 
All flagship configurations report more findings when skills are enabled.
However, precision changes only modestly.
For three of the four configurations, precision remains within $\pm$7.5\% of the without-skills baseline, while Codex/GPT-5.5 achieves a relative gain of +14.8\%.

\noindent
\textbf{Detect Award.}
Audit skills have mixed effects on Detect Award.
Antigravity/Gemini-3.5-Flash achieves the largest gain (+94.1\%) because a few high-value findings are detected, and Codex/GPT-5.5 also improves substantially (+43.2\%).
In contrast, OpenCode/GLM-5.1 declines by -25.6\% because it adds more low-value findings while losing high-value ones, while Claude/Opus-4.8 remains nearly unchanged (-2.8\%).

\begin{figure}[t]
    \centering
      \begin{minipage}{0.49\linewidth}
      \centering
      \includegraphics[width=\linewidth]{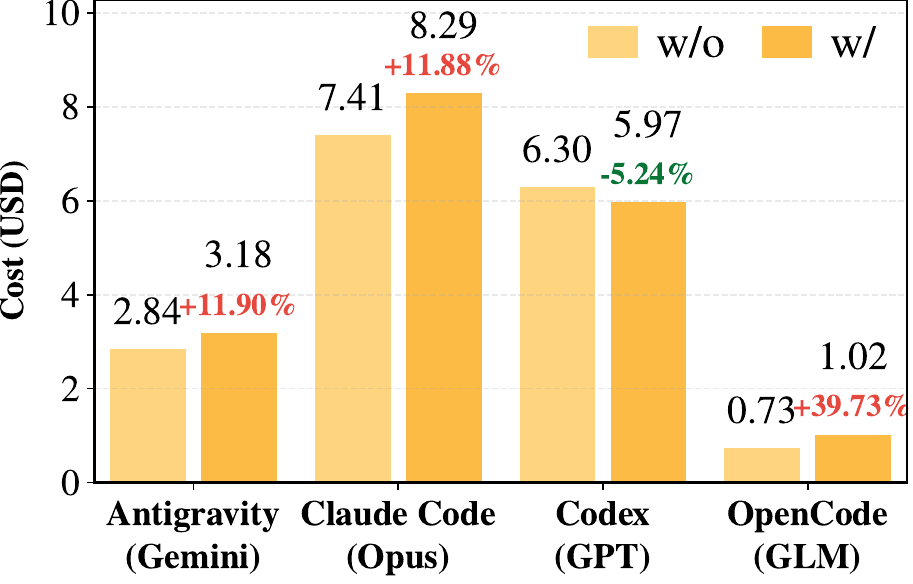}
      \subcaption{Average token cost.}
      \label{fig:avg-cost}
    \end{minipage}
    \begin{minipage}{0.49\linewidth}
        \centering
        \includegraphics[width=\linewidth]{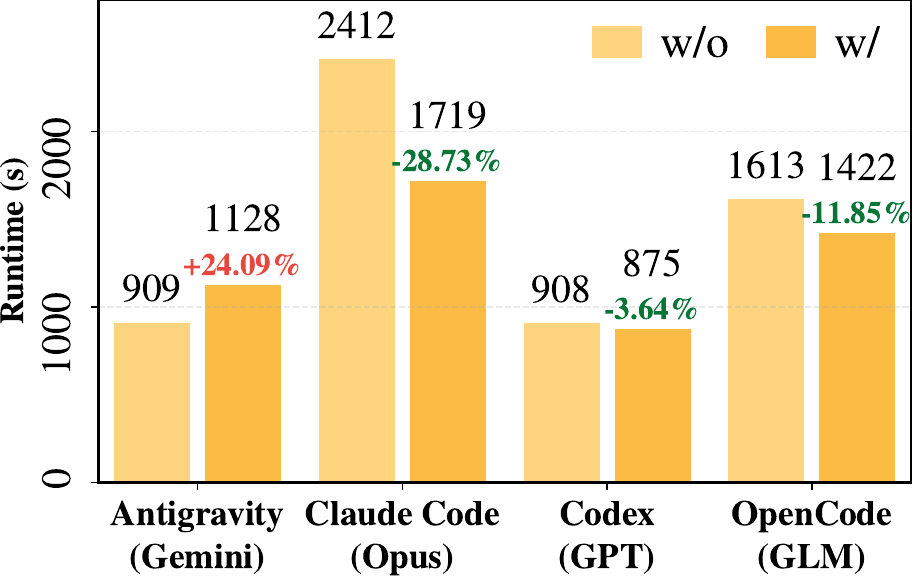}
        \subcaption{Average runtime.}
        \label{fig:avg-runtime}
    \end{minipage}
    \caption{Detection cost under the flagship model setting.}
    \label{fig:detection-cost}
\end{figure}

\noindent
\textbf{Detection Cost.}
\myfig~\ref{fig:detection-cost} shows that enabling audit skills does not introduce uniform overhead.
Overall, three of the four with-skills configurations consume more tokens, while three complete faster.
A finer-grained breakdown reveals distinct patterns across agents.
Codex/GPT-5.5 exhibits a clear \emph{Pareto improvement}~\cite{pareto2014manual}: it becomes more accurate (+22.8\%), more token-efficient ($-5.24\%$ cost per run), and faster ($-3.64\%$ runtime).
In contrast, Antigravity/Gemini-3.5-Flash is the only configuration that becomes both slower ($+24.09\%$ runtime) and more expensive ($+11.90\%$).
The remaining two show mixed behavior: token usage increases (Claude/Opus-4.8 $+11.88\%$, OpenCode/GLM-5.1 $+39.73\%$), while runtime decreases (Claude/Opus-4.8 $-28.73\%$, OpenCode/GLM-5.1 $-11.85\%$).
This suggests that skill invocation changes execution patterns rather than uniformly increasing cost.

\begin{tcolorbox}[size=title]
\noindent\textbf{Finding 6.}
Under flagship models, audit skills consistently improve Detect Score (3.2\%--22.8\% relative gains), while their effects on Disclosure Precision and Award are more limited and configuration-dependent.
They also introduce non-uniform computational overhead.
Codex/GPT-5.5 is the strongest case, improving detection effectiveness while reducing both token cost and runtime.
\end{tcolorbox}
\subsection{Ablation Study under the Same Weaker Backend Model}
\label{sec:rq2-ablation}

To better understand the uneven effectiveness of audit skills under native flagship configurations in \mysec\ref{subsec:rq2-results}, we conduct a controlled ablation to separate the effect of backend model capability from that of agent harness design.
As shown in the same-backend setting of \mytab~\ref{tab:agent-configurations-effectiveness}, we use the weaker yet cost-effective DeepSeek-V4-Pro as a unified backend for the three evaluated harnesses that support configurable model backends.

\noindent
\textbf{Model Factor.}
Compared with the native flagship setting, the weaker backend model setting substantially weakens the effect of audit skills.
Across the three harnesses, the observed skill effects become negative, ranging from -5.9\% to -9.4\%.
This suggests that backend model capability is a key determinant of whether agents can effectively leverage audit skills.

\noindent
\textbf{Agent Factor.}
Under the fixed DeepSeek-V4-Pro backend, different agent harnesses exhibit similar effectiveness.
The with-skills condition also produces a consistent downward trend across harnesses, with variations generally within 11\%.
This suggests that, once model capability is controlled, harness design has limited impact on audit-skill effectiveness.

\begin{figure}[t]
    \centering
      \begin{minipage}{0.49\linewidth}
      \centering
      \includegraphics[width=\linewidth]{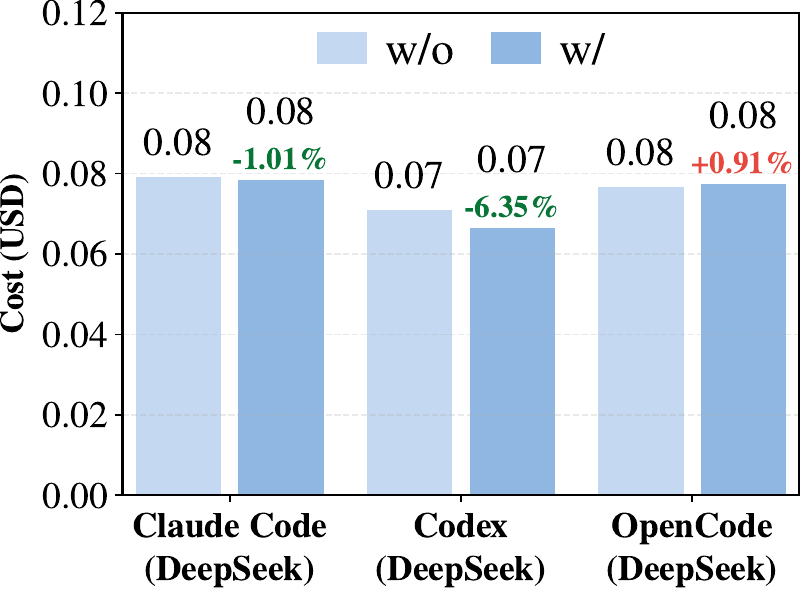}
      \subcaption{Average token cost.}
      \label{fig:avg-cost-ds}
    \end{minipage}
    \begin{minipage}{0.49\linewidth}
        \centering
        \includegraphics[width=\linewidth]{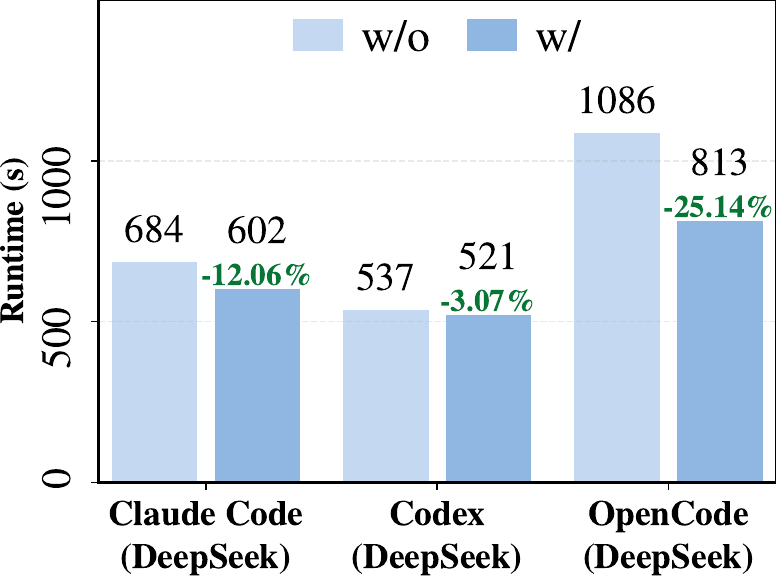}
        \subcaption{Average runtime.}
        \label{fig:avg-runtime-ds}
    \end{minipage}
    \caption{Detection cost under the DeepSeek-V4-Pro setting.}
    \vspace{-3mm}
    \label{fig:deepseek-cost} 
\end{figure}

\noindent
\textbf{Detection Cost.}
As shown in \myfig~\ref{fig:deepseek-cost}, audit skills introduce only mild computational overhead under the DeepSeek-V4-Pro setting.
Token cost varies within a narrow range from $-6.35\%$ to $+0.91\%$, while runtime consistently decreases across harnesses, ranging from $-3.07\%$ to $-25.14\%$.

\begin{tcolorbox}[size=title]
\noindent\textbf{Finding 7.} 
Under the weaker backend model, audit-skill effectiveness drops consistently across harnesses, indicating that backend model dominates harness design.
\end{tcolorbox}


\noindent
\underline{\textbf{RQ2 Summary:}} Audit-skill effectiveness is driven primarily by backend model capability rather than agent harness design.
Under strong flagship models such as Codex/GPT-5.5, audit skills can substantially improve detection effectiveness with no extra computational overhead.
In contrast, under the weaker DeepSeek-V4-Pro backend, their benefits become marginal or negative.
These results suggest that audit skills have limited standalone effectiveness when the underlying model is not capable enough to discover, interpret, and apply them.






\section{RQ3: Behavioral Impact}
\label{sec:behavior}

To analyze how audit skills affect agent behavior, we study their impact at three levels.
First, we test whether skill availability induces statistically significant shifts in execution trajectories.
Second, we measure skill triggering to determine whether agents autonomously load and invoke skills during execution.
Finally, we examine trajectory-level changes in workflows and decision-making after skills become available.

\subsection{Skill-Induced Trajectory Shifts}
\label{subsec:rq3-significance}

\begin{figure}[t]
	\centering
	\includegraphics[width=\linewidth]{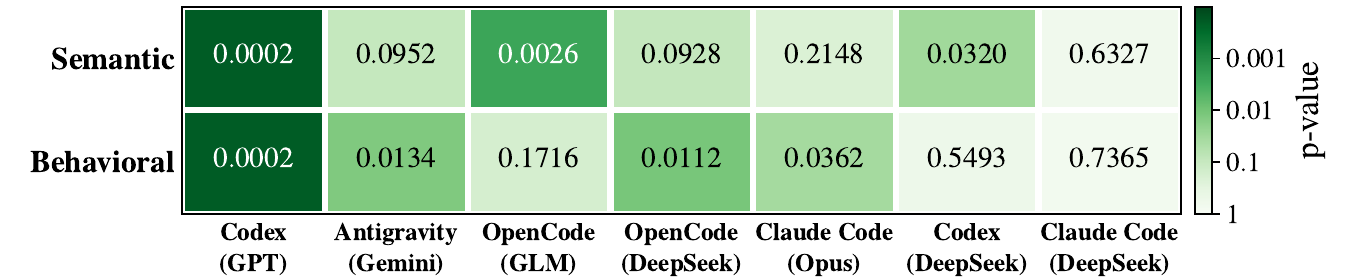}
	\caption{Significance of skill-induced trajectory shifts.}
	\vspace{-3mm}
	\label{fig:sig-heatmap}
\end{figure}

We evaluate trajectory shifts from two perspectives.
\ding{172} The \emph{semantic view} captures changes in reasoning content by embedding each trajectory with \texttt{text-embedding-3-large}, followed by length-normalized pooling and PCA-based dimensionality reduction for stable covariance estimation.
\ding{173} The \emph{behavioral view} represents each trajectory as a 10-dimensional action-type share vector, where each dimension is the proportion of a specific action type (reason, skill, read, list, grep, edit, bash, plan, search, or other) in the action sequence.
This view isolates structural behavioral changes from textual semantics.

For each view, we compute per-task paired differences between with-skill and without-skill executions and test for zero mean shift using a paired Hotelling's $T^2$ statistic~\cite{hotelling1931generalization}.
Given the moderate sample size, we use a sign-flip permutation test ($B{=}5000$) to obtain empirical $p$-values for each configuration.
These $p$-values quantify whether the observed trajectory shifts are statistically significant.

As shown in \myfig~\ref{fig:sig-heatmap}, most agent configurations exhibit statistically significant trajectory shifts ($p < 0.05$) in either the semantic or behavioral view.
The only exception is Claude/DeepSeek, which remains non-significant in both views.
Codex/GPT shows the strongest and most consistent effect, with $p < 0.001$ in both the semantic ($p=0.0002$) and behavioral ($p=0.0002$) views.
This indicates that audit skills can reshape both reasoning content and execution actions.

\begin{tcolorbox}[size=title]
\noindent\textbf{Finding 8.} 
Six of seven agent configurations exhibit statistically significant trajectory shifts in at least one of the semantic or behavioral views, indicating that audit skills induce measurable behavioral changes in most settings.
\end{tcolorbox}

\subsection{Skill Triggering}
\label{subsec:rq3-activation}

In the RQ2 \textit{with-skills} setting, all skills are available to the agent, but only their metadata (i.e., name and description) is exposed in the initial context as lightweight discovery signals.
A skill's full instruction body is loaded into the working context only when the agent explicitly invokes it through a tool call.
We therefore measure skill triggering from session traces and count a skill as activated only when its full content is explicitly read (e.g., loading \texttt{SKILL.md}).

As shown in \mytab~\ref{tab:skill-triggering-results}, we report, for each setup, the number of audits with triggered skills, the number of distinct skills activated, the total number of skill reads, and the corresponding changes in detection effectiveness from \mytab~\ref{tab:agent-configurations-effectiveness}.

\noindent
\textbf{Trigger Volume.}
Skill triggering varies substantially across setups, ranging from no activation to activation in all 40 audits.
The number of distinct skills actually used is also limited: at most 27 of the 83 available skills are invoked, and total skill reads range from 0 to 119.
Overall, only Codex/GPT-5.5 shows consistently high activation, while all other configurations trigger skills only sporadically or not at all.

\noindent
\textbf{Determinants of Activation.}
Controlled comparisons indicate that model capability is the primary driver of skill triggering.
(i) Under the same Codex CLI and identical skill exposure, GPT-5.5 triggers skills in all 40 audits, whereas DeepSeek-V4-Pro triggers none, suggesting that instruction-following capability determines whether available skills are used.
(ii) When DeepSeek-V4-Pro is used across different agent harnesses, skill activation remains near zero, indicating that harness choice has limited impact on triggering under a weaker model.

\noindent\textbf{Relation to Effectiveness.}
Skill activation is closely aligned with detection effectiveness.
The only setup with universal activation (Codex/GPT-5.5) is also the only one with large detection gains, whereas setups with sparse or no activation show negligible or negative improvements.
This suggests that limited improvement in most settings is driven primarily by insufficient skill utilization rather than by a lack of skill utility.

\begin{tcolorbox}[size=title]
\noindent\textbf{Finding 9.} 
Skill triggering is a primary bottleneck to its effectiveness: only one setup activates skills consistently across audits.
Triggering is driven mainly by the model itself, while the agent harness has limited impact.
\end{tcolorbox}
\subsection{Trajectory Impact of Skill Invocation}
\label{subsec:rq3-workflow}

Because skill invocation is implemented through tool calls, we represent each execution trajectory as an ordered timeline of tool-call events extracted from the structured session trace.
It preserves action order, allowing us to localize skill activation and analyze how subsequent tool-call sequences evolve.

\begin{table}[t]
\centering
\caption{Skill triggering per setup (\% change from Table~\ref{tab:agent-configurations-effectiveness}).}
\label{tab:skill-triggering-results}
\setlength{\tabcolsep}{1pt}
\footnotesize
\resizebox{\linewidth}{!}{
\begin{tabular}{lccccccc}
\toprule
\textbf{Agent / Model} & \textbf{\makecell{Activ.\\Audits}} & \textbf{\makecell{Dist.\\Skills}} & \textbf{\makecell{Total\\Reads}} & \textbf{\makecell{Score\\\%}} & \textbf{\makecell{Prec.\\\%}} & \textbf{\makecell{Award\\\%}} \\
\midrule
Codex / GPT-5.5            
& \textbf{40/40} 
& \textbf{27} 
& \textbf{119} 
& \textbf{+22.8}
& \textbf{+14.8}
& +43.2
\\
Antigravity / Gemini-3.5-Flash 
& 13/40 
& 2  
& 31  
& +6.4
& $-4.3$
& \textbf{+94.1}
\\
OpenCode / GLM-5.1         
& 3/40  
& 6  
& 7   
& +13.9
& +7.4
& $-25.6$
\\
OpenCode / DeepSeek-V4-Pro 
& 1/40  
& 2  
& 2   
& $-8.8$
& $-3.3$
& $-93.1$
\\
Claude / Opus-4.8          
& 0/40  
& 0  
& 0   
& +3.2
& $-6.5$
& $-2.8$
\\
Codex / DeepSeek-V4-Pro    
& 0/40  
& 0  
& 0   
& $-5.9$
& $-0.4$
& $-10.9$
\\
Claude / DeepSeek-V4-Pro   
& 0/40  
& 0  
& 0   
& $-9.4$
& $-10.2$
& $-0.8$
\\
\bottomrule
\end{tabular}
}
\end{table}

\begin{figure}[t]
	\centering
	\includegraphics[width=\linewidth]{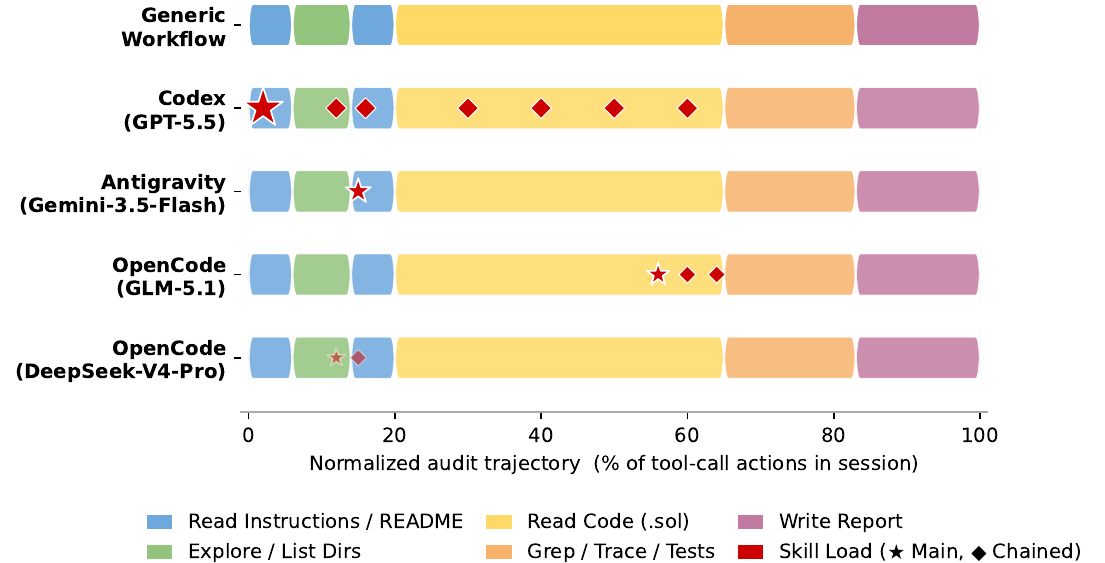}
	\caption{Positions of skill-load events along the audit trajectory.}
	\label{fig:skill-trigger-timeline}
\end{figure}

\noindent
\textbf{Skill Invocation Patterns.}
\myfig~\ref{fig:skill-trigger-timeline} locates skill-load events along the audit timeline.
To compare setups with different execution lengths, we normalize each trajectory by the fraction of elapsed tool calls.
Across model configurations, audits share a common six-stage backbone: instruction reading $\rightarrow$ directory exploration $\rightarrow$ scoped \texttt{README} inspection $\rightarrow$ code analysis $\rightarrow$ grep/trace/testing $\rightarrow$ report generation.
Within this shared workflow, skill invocation differs along two dimensions:

\ding{172} \emph{Activation position.} Configurations differ systematically in when skills are introduced into the workflow. Codex/GPT-5.5, OpenCode/DeepSeek-V4-Pro, and Antigravity/Gemini-3.5-Flash tend to load skills early, typically during the pre-analysis stage before intensive code inspection begins. In contrast, OpenCode/GLM-5.1 triggers skills only after substantial code exploration, indicating delayed use of external guidance.

\ding{173} \emph{Invocation pattern.} We observe two modes of skill usage.
In the \emph{single} mode, agents rely on one primary workflow skill throughout the task, as seen in Antigravity/Gemini-3.5-Flash and OpenCode/DeepSeek-V4-Pro.
In the \emph{chained} mode, agents dynamically load multiple skills across audit stages, adapting skill usage to the task context.
A representative example from the 2024-03-taiko audit is:
\texttt{blockchain-auditor}\,$\rightarrow$\,\texttt{bridges}\,$\rightarrow$\,\texttt{erc20}\,$\rightarrow$\, \texttt{precision-math}\,$\rightarrow$\,\texttt{proxies}, where each skill is invoked in response to the code region that triggers it.
This context-aligned, stage-wise composition appears in Codex/GPT-5.5 and OpenCode/GLM-5.1, reflecting more adaptive orchestration of domain knowledge across the execution trajectory.

\begin{table}[t]
\centering
\caption{Trajectory changes induced by skill activation.}
\label{tab:skill-design-footprint}
\setlength{\tabcolsep}{1pt}
\footnotesize
\begin{tabular}{p{4cm}lc}
\toprule
\textbf{Agent / Model} & \textbf{Trajectory Changes} & \textbf{\#Audit Tasks} \\
\midrule
\multirow{6}{=}{Codex / GPT-5.5}
 & K1: Checklist        
 & 15 \\
 & K2: Incident          
 & 7  \\
 & T1: Slither
 & 18 \\
 & T2: Aderyn
 & 15 \\
 & T3: Semgrep            
 & 5  \\
 & W1: \texttt{.vigilo}            
 & 17 \\
\midrule
\multirow{2}{=}{Antigravity / Gemini-3.5-Flash}
 & T1: Slither           & 4  \\
 & T2: Aderyn            & 2  \\
\midrule
\multirow{1}{=}{OpenCode / GLM-5.1}        & ---  & --- \\
\midrule
\multirow{1}{=}{OpenCode / DeepSeek-V4-Pro} & --- & --- \\
\bottomrule
\end{tabular}
\end{table}

\noindent
\textbf{Trajectory Footprints Induced by Skills.}
After identifying skill activation points, we inspect all 57 skill-activated audits and compare each with its paired without-skill execution to characterize the structural footprints induced by skills.

As shown in \mytab~\ref{tab:skill-design-footprint}, we map these footprints to the three skill design dimensions identified in \mysec\ref{sec:design}: \emph{knowledge representation} (RQ1.2), \emph{tool dependencies} (RQ1.3), and \emph{workflow structure} (RQ1.4).
Overall, Codex/GPT-5.5 exhibits a comprehensive footprint across all three dimensions.
In the knowledge dimension, it applies structured checklists (K1) and incident-based (K2) reasoning patterns.
In the tool dimension, it substantially expands the use of analyzers such as Slither (T1), Aderyn (T2), and Semgrep (T3).
In the workflow dimension, it instantiates skill-prescribed staged and severity-bucketed scaffolds (e.g., \texttt{.vigilo}).
In contrast, Antigravity/Gemini-3.5-Flash exhibits a partial footprint concentrated mainly in the tool dimension, with increased use of Slither and Aderyn but limited changes in knowledge representation or workflow structure.
The two OpenCode configurations show no consistent footprint across any dimension, indicating minimal observable propagation of skill design into execution trajectories.

\begin{tcolorbox}[size=title]
\noindent\textbf{Finding 10.} 
Skill invocation preserves the core audit workflow backbone but introduces distinct activation positions and single or chained usage patterns across audit stages.
Its downstream footprint varies across agent--model configurations, affecting knowledge usage, tool invocation, and workflow structure to different degrees.
\end{tcolorbox}

\section{Discussion}

\subsection{Robustness Analysis}

Given the inherent stochasticity of LLMs, we repeat the Codex/GPT-5.5 experiment three times under identical settings.
As shown in \mytab~\ref{tab:codex-three-runs}, the with-skills configuration consistently improves Detect Score and Award across all runs, while Precision remains relatively stable with only minor fluctuations.
The direction of improvement is preserved in every run, suggesting that the observed gains are not artifacts of random variation but a stable effect of skill augmentation.

\subsection{Threats to Validity}

\noindent
\textbf{Internal Validity.}
Most of our results are based on single-run executions (due to the LLM agent monetary cost), which may be affected by LLM stochasticity.
We mitigate this concern with the robustness analysis above, which shows stable directional improvements across repeated Codex/GPT-5.5 runs.
In addition, we evaluate skills under autonomous invocation, which reflects realistic deployment but does not isolate the effect of each individual skill.
Designing standardized benchmarks for per-skill causal attribution is left to future work.

\noindent
\textbf{External Validity.}
Our findings may not generalize to all benchmarks, models, or skill ecosystems.
(i) Although EVMBench may raise potential data leak concerns, our study focuses on the \emph{relative effect} of skill augmentation rather than absolute performance.
(ii) Due to monetary constraints, we evaluate a representative set of frontier and cost-effective models rather than exhaustively covering all available agents.
(iii) Our skill corpus is a temporal snapshot of a rapidly evolving ecosystem.
To support reproducibility and future extension, we release the curated data and evaluation scripts.

\begin{table}[t]
\centering
\caption{Results of Codex/GPT-5.5 across 3 repeated runs.}
\label{tab:codex-three-runs}
\footnotesize
\setlength{\tabcolsep}{1pt}
\resizebox{\linewidth}{!}{
\begin{tabular}{llccc}
\toprule
\textbf{Run} & \textbf{Skills} & \textbf{Detect Score} & \textbf{Disclosure Precision} & \textbf{Detect Award} \\
\midrule
R1 
& w/o 
& 79/120 (65.8\%) 
& 79/274 = 28.8\% 
& 49.7\% 
\\
& w/    
& 97/120 (80.8\%) \textcolor{green!70!black}{(+22.8\%)}
& 97/293 = 33.1\% \textcolor{green!70!black}{(+14.8\%)}
& 71.2\% \textcolor{green!70!black}{(+43.2\%)}
\\
\midrule
R2 
& w/o 
& 79/120 (65.8\%) 
& 79/242 = 32.6\% 
& 64.3\% 
\\
& w/    
& 95/120 (79.2\%) \textcolor{green!70!black}{(+20.3\%)}
& 95/284 = 33.5\% \textcolor{green!70!black}{(+2.5\%)}
& 74.6\% \textcolor{green!70!black}{(+16.0\%)}
\\
\midrule
R3 
& w/o 
& 78/120 (65.0\%) 
& 78/226 = 34.5\% 
&  56.1\% 
\\
& w/    
& 100/120 (83.3\%) \textcolor{green!70!black}{(+28.2\%)}
& 100/324 = 30.9\% \textcolor{red!60!black}{(-10.6\%)}
& 73.2\% \textcolor{green!70!black}{(+30.5\%)}
\\
\bottomrule
\end{tabular}
}
\end{table}

\section{Related Work}
\label{sec:related}

Since the emergence of agent skills, 
research has begun to examine this paradigm from multiple perspectives.
For \emph{design}, 
SkillReducer~\cite{gao2026skillreducer} identifies systemic inefficiencies and compresses skills for token efficiency.
For \emph{generation}, 
Auto-SKILL.md~\cite{hao2026automating} mines interaction trajectories into skill documents, 
OpenClaw-Skill~\cite{lin2026openclaw} searches a collective skill tree across models, 
and SkillNet~\cite{skillnet} provides open infrastructure for creating and connecting skills at scale.
For \emph{composition}, SkillWeaver~\cite{gao2026compositional} uses decompose-retrieve-compose routing, 
while SkillOpt~\cite{yang2026skillopt} performs \emph{optimization} through a transferable text-space optimizer.
SkillsBench~\cite{Li2026SkillsBenchBH} supports paired \emph{evaluation} of skill efficacy, 
SkillMAS~\cite{pan2026skillmas} studies \emph{application} by co-evolving skills with multi-agent systems, 
and a \emph{security} study~\cite{liu2026agent} reports that 26.1\% of skills contain vulnerabilities.

However, these efforts largely study agent skills in a domain-agnostic manner, often leaving the analysis at a black-box level with no in-depth, domain-specific insights.
In parallel, recent advances in LLMs have improved smart contract security tasks, including auditing, repair, and exploitation. 
A growing line of LLM-based methods, 
such as GPTScan~\cite{sun2024gptscan}, PropertyGPT~\cite{Liu2024PropertyGPTLF}, iAudit~\cite{ma2025combining}, and LLM-SmartAudit~\cite{Wei2025AdvancedSC}, 
has demonstrated strong auditing capabilities, 
and benchmarks such as EVMBench~\cite{Wang2026EVMbenchEA} have begun to assess agent-based auditing. 
More recently, a growing number of smart contract audit skills have been developed to augment these agents.
Yet little is known about how such skills are designed, 
how effective they are in practice, 
or how they influence agent behavior.
Our work addresses this gap through the first systematic empirical study of smart contract audit skills.








\section{Conclusion}
\label{sec:conclusion}

In this paper, we presented an empirical study of smart contract audit skills, curating 83 skills and evaluating their design, effectiveness, and behavioral impact across seven agent--model configurations on EVMBench.
We find that audit skills are lightweight yet heterogeneous in design, that their benefits depend mainly on backend model capability, and that skill triggering is the key bottleneck to their behavioral impact.
When invoked, skills preserve the core audit workflow. 

\bibliographystyle{IEEEtranS}
\bibliography{main}

\end{document}